\documentclass[fleqn]{article}

\usepackage{aaai}
\nocopyright

\usepackage{url}
\usepackage{epsfig}
\usepackage{amsmath}
\usepackage{latexsym}
\usepackage{program}

\newcommand{\la}{\leftarrow}
\newcommand{\La}{\Leftarrow}
\newcommand{\ra}{\rightarrow}
\newcommand{\lra}{\leftrightarrow}
\newcommand{\ind}{\qquad\tab}
\newcommand{\ol}[1]{\overline{#1}}
\newcommand{\st}[2]{\langle#1\;\origbar\origbar\;#2\rangle}
\renewcommand{\C}[1]{\mathcal{#1}}
\newcommand{\V}[1]{\forall #1.\;\:}
\newcommand{\E}[1]{\exists #1.\;\:}
\newcommand{\Not}{not\;}
\newcommand{\smodels}{\textit{smodels}}
\newcommand{\datalog}{\textsc{datalog}}
\newcommand{\datalogc}{\textsc{datalog}$^{CIRC}$}
\newcommand{\npspec}{\textsc{np-spec}}

\begin{document}

\title{A Comparison of Logic Programming Approaches for Representation
  and Solving of Constraint Satisfaction Problems}
\author{Nikolay Pelov \and Emmanuel De Mot \and Maurice Bruynooghe \\[10pt]
  Department of Computer Science, K.U.Leuven \\
  Celestijnenlaan 200A, B-3001 Heverlee, Belgium \\
  E-mail: {\tt $\{$pelov,emmanuel,maurice$\}$@cs.kuleuven.ac.be} 
}

\maketitle

\begin{abstract}
Many logic programming based approaches can be used to describe and
solve combinatorial search problems. On the one hand there are
definite programs and constraint logic programs that compute a
solution as an answer substitution to a query containing the
variables of the constraint satisfaction problem. On the other hand
there are approaches based on stable model semantics, abduction, and
first-order logic model generation that compute solutions as models
of some theory. This paper compares these different approaches from
point of view of knowledge representation (how declarative are the
programs) and from point of view of performance (how good are they
at solving typical problems).

\end{abstract}

\section{Introduction}

Consistency techniques are widely used for solving finite constraint
satisfaction problems (CSP). These techniques have been integrated in
logic programming, resulting in finite domain constraint logic
programming \cite{clp:cslp}. In this paradigm, a program typically
creates a data structure holding the variables of the CSP to be
solved, sets up the constraints and uses a labelling technique to
assign values to the variables. The constraint solver uses consistency
techniques \cite{clp:Tsang} to prune the search. Solutions are given
by an answer substitution to a goal. This leads to a rather procedural
programming style. Moreover, the problem description is not very
declarative because the mapping between domain variables and their
value has an indirect representation in a term structure. A more
detailed discussion for representing and solving CSP in different
logical systems, following this approach, can be found in
\cite{clp:Mackw92}.

An alternative representation of a CSP can be given in a first-order
logic setting, where instead of using variables, one can define the
problem in a more natural way by using constant and function symbols
and specify the constraints as logic formulae. A solution is then given
by a model of the theory, and in particular by the interpretation of
the function symbols. We argue that representing a problem in this way
tends to be more declarative. To achieve a similar declarative
representation in a logic programming system, the functions should be
replaced with predicates and then an answer of a CSP is given by a
table of facts. Abduction \cite{abd:ALP} is one framework which allows
such reasoning. The relation between the variables of the CSP and
their values is declared as an abducible and solution of the problem
is given by a set of abduced atoms. More recently, a logic programming
paradigm based on stable model semantics \cite{sem:SMS} has emerged
and Niemel{\"a} \cite{mg:smodels} proposes it as a constraint solving
paradigm. In this approach a solution is given by a stable model of
the program. 

In this paper, we use two typical CSP problems to compare the merits
of the various approaches. One is graph coloring where the domain
size is constant but the number of constraints increases with
increasing problem size; the other is the n-queens problem where both
the domain size and the number of constraints increases with
increasing problem size. For both problems we show the representation
in the various approaches\footnote{We do not use the actual syntax of
  the systems but a syntax which is more uniform across the different
  systems.}, comment on the declarativity of the representation and
briefly describe the basic principles behind the implementation. We
also compare the performance of different systems and show how it
scales by increasing the problem size.

First, we give some basic notions about constraint satisfaction and
logic programming. Then we discuss the classical representation of CSP
as a logic and constraint logic program.
The next section gives a different representation of a CSP and shows
how it can be translated to an equivalent logic program which then can
be solved by abductive reasoning or by computing stable models of
the program. Finally, we give the results of solving some well-known
problems with several systems and the last section gives a summary of
our research and some directions for future research.

\section{Preliminaries}

A {\em constraint satisfaction problem} is usually defined as a finite
set of {\em constraint variables} $\C{X} = \{X_1,\ldots,X_n\}$, a
finite domain $D_i$ of possible values for each variable $X_i$, and a
finite set of {\em constraint relations}
$\C{R}=\{r_{S_1},\ldots,r_{S_t}\}$, where $S_i\subseteq \{1,\ldots,n\}$
for $i=1,\rdots,t$ are the indices of the variables which are
constrained by $r_{S_i}$.  A {\em solution} is an instantiation of the
variables $\C{X}$ which satisfies all the constraints in $\C{R}$.

A logic program is defined for a fixed language $\C{L}$ of function
and predicate symbols and an infinite set of variables $\C{X}$.  We
assume that our language always contains the equality predicate $=/2$.
Terms and atoms are defined as usual. A {\em literal} is either an
atom $a$ or a negation of an atom $\Not a$. A {\em clause} has the
form
\[a\la b_1,\ldots,b_n.\]
where $a$ is an atom and is called {\em head} and $b_i$ are literals
and $b_1,\ldots,b_n$ is called {\em body}. A clause is called definite
if all $b_i$ are positive.  A (definite) logic program $P$ is a set of
(definite) clauses. A Herbrand interpretation and Herbrand model are
again defined as usual (cf. \cite{lloyd}). A definite logic program
always has a unique least Herbrand model which will be denoted with
$lm(P)$.

\section{Knowledge Representation of CSP}

\subsection{Logic Programming}

A CSP can be easily represented as a definite logic program. The
domains of the variables are defined by unary predicates. A constraint
relation $r/n$ is defined by a predicate $p/n$ such that $lm(p/n) =
r/n$.  Then a CSP can be defined by a clause of the form:
\begin{program}
\ind csp(X_1,\ldots,X_n)\la 
\ind d_1(X_1),\ldots,d_n(X_n),
     p_1(X_{S_1}),\ldots, p_t(X_{S_t}).
\end{program}
A computed answer substitution $\theta$ for the goal $\la
csp(X_1,\ldots,X_n)$ using SLD resolution will be a solution to the
CSP since $P\models csp(X_1,\ldots,X_n)\theta$ and in particular
$lm(P)\models csp(X_1,\ldots,X_n)\theta$. It is also easy to see that
the least model of the $csp/n$ predicate will contain all solutions to
the problem. So, to compute all solutions to a given problem one can
use a bottom-up fix-point operator to compute the least model of a
program.

To give an example let us consider the problem for finding the
positions of $n$ queens on a $n\times n$ chess board such that no two
queens attack each other. One way of formalizing this problem is by
using $n^2$ boolean variables which indicate whether a queen is placed
on a particular square. However, by taking into account the fact that
no two queens can be placed on the same column, we can use only $n$
variables which give the positions of the queens for each column. Then
for all pairs of queens we should state the constraint that they can
not attack each other by being on a same row or diagonal. This
constraint can be defined with the following predicate which is
parameterized by the distance $D$ between the columns of the two
queens:
\begin{program}
\ind safe(X_1,X_2,D)\la X_1\neq X_2, abs(X_1-X_2)\neq D.
\end{program}
Then the 4-Queens problem can be defined as:
\begin{program}
\ind csp(X_1,X_2,X_3,X_4) \la
\ind d(X_1),d(X_2),d(X_3),d(X_4),
     safe(X_1,X_2,1),safe(X_2,X_3,1),
     safe(X_1,X_3,2),safe(X_1,X_4,3),
     safe(X_2,X_4,2),safe(X_3,X_4,1).
\end{program}
Here we exploit another particularity of the n-Queens problem, that
the $safe$ constraint is symmetric for the first two arguments.
Hence it is enough to only check queens such that the column of the
first one is less than the column of the second one.
Executing the query $\la csp(X_1,X_2,X_3,X_4)$ under the top-down
left-to-right strategy of Prolog will result in solving the problem by
a generate and test approach.
By interleaving the calls to the domain predicates $d/1$ and the
constraint predicates will result in a standard backtracking:
\begin{program}
\ind csp(X_1,X_2,X_3,X_4) \la
\ind d(X_1),d(X_2),\; safe(X_1,X_2,1),
     d(X_3),\; safe(X_2,X_3,1),safe(X_1,X_3,2),
     d(X_4),\;\tab safe(X_1,X_4,3),safe(X_2,X_4,2),
                   safe(X_3,X_4,1).
\end{program}

A lot of research has been done to improve the execution strategy of
standard Prolog. For example, a technique known as {\em co-routing}
uses a literal selection strategy which selects a
constraint as soon as all its arguments become ground. This allows a
generate and test program to be executed with standard backtracking.
Another technique known as {\em intelligent backtracking}
\cite{ib:mb-91} does a failure analysis and backtracks only to
variable assignments which are actually responsible for the failure.

\begin{figure}[htb]
\begin{program}
\ind csp(N,L)\la
\ind make\_vars(L,N),
     constrain\_all(L). \untab
~
make\_vars([],0).
make\_vars([H\origbar T],N) \la
\ind N > 0, 
     d(H),
     N1\; is\; N - 1,
     make\_vars(T,N1). \untab
~
constrain\_all([]).
constrain\_all([X\origbar Xs]) \la
\ind constrain\_all(Xs),
     constrain\_between(X,Xs,1). \untab
~
constrain\_between(X,[],N).
constrain\_between(X,[Y\origbar Ys],N) \la
\ind safe(X,Y,N),
     N_1\; is\; N+1,
     constrain\_between(X,Ys,N_1).
\end{program}
\caption{N-Queens as a Definite Logic Program} \label{fig:queen}
\end{figure}

An important problem with this representation of a CSP is that it has
to be defined specifically for a given number of queens. If we want to
parameterize the problem with respect to the number of queens then we
should use some data structure (usually a list) to store the
constraint variables. Figure \ref{fig:queen} shows a typical
specification of the problem. The $make\_vars/2$ predicate constructs
a list of $n$ variables and uses backtracking to enumerate all
possible solutions.

However this way of representing the problem is even less declarative
than before. First of all, the column number of each queen is
implicitly defined by the position of the corresponding variable in
the list. Then, in order to add constraints only between queens
such that the column of the first one is less than the column of the
second one we have to use two nested recursive predicates.

\begin{figure}[htb]
\begin{program}
\ind graph\_coloring(Vert,Edges,Vars)\la
\ind	make\_vars(Vert,Vars),
	add\_constr(Edges,Vars). \untab
~
make\_vars([],[]).
make\_vars([V\origbar Vert],[assoc(V,X)\origbar Vars])\la
\ind	color(X),
	make\_vars(Vert,Vars). \untab
~
add\_constr([],Vars).
add\_constr([edge(V_1,V_2)\origbar Edges],Vars)\la
\ind	member(assoc(V_1,X_1),Vars),
	member(assoc(V_2,X_2),Vars),
	X_1 \neq X_2,
        add\_constr(Edges,Vars).
\end{program}
\caption{Graph Coloring as a Definite Logic Program} \label{fig:gcp}
\end{figure}

Another typical CSP problem is that of graph coloring.  The goal is to
color all the vertices of a graph in such a way that no two adjacent
vertices have the same color.  The graph can be represented by a list
of vertices and a list of terms $edge(v_1,v_2)$ describing the edges.
The problem can be expressed as a CSP by associating a different
variable for each vertex and restricting its domain to all possible
colors. Then for each edge in the graph we put a disequality
constraint between the variables corresponding to the vertices of the
edge. Figure \ref{fig:gcp} shows a sample formulation of this problem
as a logic program. Here the correspondence between a vertex and the
color assigned to it is made explicit by means of an association list.

\subsection{Constraint Logic Programming}

Constraint logic programming (CLP) \cite{clp:survey@JLP} is an
extension of logic programming where some of the predicate and
function symbols have a fixed interpretation. This interpretation is
dependent on a particular constraint domain $X$ (e.g. finite trees or
real numbers) and this allows for a much more natural representation
of problems from this particular domain.  Besides its better
declarative semantics for expressing problems, an implementation of a
CLP($X$) system also includes an efficient domain specific solver
$solve_X$ for checking satisfiability of a set of constraints.

A proof procedure for CLP is defined as an extension of standard
resolution. A state is defined as a pair $\st{\la a,A}{C}$ of a goal
and a set of constraints. At each step of the computation, some
literal $a$ is selected from the current goal according to some
selection function. If $a$ is a constraint predicate then the next
state is $\st{\la A}{C\wedge a}$ if $solve(C\wedge a)\neq false$ or
$\st{\Box}{false}$ otherwise. If $a$ is a normal atom then the next
state is $\st{\la\vec{s}=\vec{t},B,A}{C}$ for some clause $b\la B$ where
$a$ is of the form $p(\vec{s})$ and $b$ is of the form $p(\vec{t})$.

A well suited constraint domain for representing CSPs is that
of finite domain integer arithmetic $CLP(FD)$
\cite{clp:clp(FD),clp:cc(FD)}. It includes standard arithmetic
relations like $=,\neq,<$ and functions $+,-,*$ with their usual
interpretation. The implementation of $CLP(FD)$ is based on
consistency  \cite{clp:Tsang} algorithms.

A CSP represented as a logic program can be translated to a CLP(FD)
program in a straightforward way. First, the domains of the variables
are declared by a special CLP predicate (for example $X\; in\; 1..n$)
and the constraints are defined using the CLP predicates and
functions.  Then the execution of the goal $\la csp(X_1,\ldots,X_n)$
by a constraint proof procedure will result in a set of constraints
which are then solved efficiently by consistency techniques.

\section{An Alternative Representation}

From the examples and the discussion in the previous section, the
following general methodology can be given for representing a
parameterized CSP as a (constraint) logic program. First, we create
some data structure where each value of the parameter of the
problem (e.g. the number of queens, or the number of vertices) is
associated with a different variable. Then we define some recursive
predicates which iterate over this data structure and define
constraints between the variables whose corresponding parameters
satisfy certain conditions.

It seems natural to represent the mapping between parameter and
variable with a function. First let us introduce a new domain $D_p$
which includes all values of the parameter. For example, for the
n-Queens problem $D_p$ is the set of columns and for the graph
coloring problem $D_p$ will contain all vertices. Then a solution to a
CSP can be represented by a function $f: D_p\ra D$, where $D$ is the
domain of the constraint variables.

The use of functions for representing CSPs can be realized in a
first-order logic setting where a solution of a problem can be given
by an interpretation of a function symbol in a model of the theory
describing the problem. Also the constraints are expressed more
naturally by using function symbols. A definition of the n-Queens
problem in first-order logic could be the following:

\begin{program}
\ind \V{C_1,C_2} C_1<C_2 \ra
\ind safe(pos(C_1),pos(C_2),C_2-C_1).\untab
~
\V{X_1,X_2,D} safe(X_1,X_2,D)\lra
\ind X_1\neq X_2, abs(X_1-X_2)\neq D.
\end{program}

The function symbol $pos/1$ represents the mapping from columns to
rows. A model\footnote{We assume here that the arithmetic relations
  and functions have a fixed interpretation, like in CLP.} of this
theory based on a domain with $n$ elements will consist of an
interpretation of the function $pos/1$ and will be a solution of the
problem with $n$ queens.

If we want to work with several domains with different sizes then we
may use a many sorted first-order logic where the arguments of the
function and predicate symbols are assigned (possibly different)
sorts. Then in an interpretation of our theory for each sort we can
assign a domain with the appropriate size. The full specification of
the graph coloring problem in many sorted logic is given below.

\begin{program}
\ind S = \{s_v,s_c\}
F = \{col:s_v\ra s_c\}
~
\V{V_1,V_2} col(V_1)\neq col(V_2)\la edge(V_1,V_2).
~
edge(1,2).
\ldots
\end{program}

Here we use two sorts - $s_v$ for vertices, and $s_c$ for colors and a
function $col:s_v\ra s_c$ which maps vertices to colors. The edges of
the graph are described as a set of facts $edge(v_1,v_2)$. An
interpretation of this theory should associate the set with all
vertices of the graph with the sort $s_v$ and a set with possible
colors with the sort $s_c$.

Several systems for generation of finite models of many sorted
first-order logic theories are available (e.g.\ FINDER
\cite{mg:finder} and SEM \cite{mg:SEM}) and thus can be used for
solving CSPs.

\subsection{Back to Logic Programming}

Using functions for representing CSPs in a logic programming setting
is not possible. The reason is that the domain of the computation is
the Herbrand domain (or comes from the constraint domain) and that the
interpretation of the function symbols is fixed. To overcome this
restriction, one can introduce unary predicates $d_p/1$ and $d/1$
defining the domain and the range of the functions and encode a unary
function $f(X)=Y$ as a binary predicate $p_f(X,Y)$ with domain defined
by $d_p/1$ and range by $d/1$. The following axioms establish that the
interpretation of $p_f/2$ corresponds to a function:
\begin{align}
 \V{X}\E{Y}& d(Y)\wedge p_f(X,Y)\la d_p(X). \label{eq:pred-fol} \\
 \V{X,Y}& d_p(X)\la p_f(X,Y). \label{eq:pred-nout} \\
 \V{X,Y,Z}& Y = Z\la p_f(X,Y),p_f(X,Z). \label{eq:pred-func}
\end{align}
The formula \eqref{eq:pred-fol} states that the predicate $p_f(X,Y)$
is defined for all $X$ in the domain $d_p(X)$ and that the value of
$Y$ is in the range $d(Y)$. The formula \eqref{eq:pred-nout} enforces
that the predicate $p_f$ is false for all values of $X$ not in the
domain $d_p(X)$ and \eqref{eq:pred-func} ensures that the predicate
$p_f(X,Y)$ defines a function from $X$ to $Y$. By introducing an
auxiliary predicate $has\_p$, \eqref{eq:pred-fol} can be rewritten as:
\begin{align}
\label{eq:aux1} \tag{\ref{eq:pred-fol}a}
& \V{X} has\_p(X) \lra\E{Y} d(Y)\wedge p_f(X,Y). \\
\tag{\ref{eq:pred-fol}b}
& \V{X} has\_p(X) \la d_p(X). \label{eq:aux2}
\end{align}

Formula \ref{eq:aux1} is the completion of the predicate $has\_p$.
Writing it as a logic program clause, the only-if part can be dropped.
However, the other three axioms \eqref{eq:aux2}, \eqref{eq:pred-nout},
and \eqref{eq:pred-func} do not define any predicates but are just
formulae which should be true in models of the program. In logic
programming, such clauses are known as {\em integrity constraints} and
we will denote them with the symbol $\La$\footnote{The standard form
  to write integrity constraints is as clauses with an empty head.
Here we use another notation for better clarity.}.
Figure \ref{fig:pred-func} shows the resulting logic program.

\begin{figure}[hbt]
\begin{align*}
 has\_p(X)&\la d(Y),p_f(X,Y). &&\qquad \text{from \eqref{eq:aux1}} \\[10pt]
 has\_p(X)&\La d_p(X).        &&\qquad \text{from \eqref{eq:aux2}} \\
 d_p(X)&\La p_f(X,Y).         &&\qquad \text{from \eqref{eq:pred-nout}} \\
 Y=Z &\La p_f(X,Y),p_f(X,Z).  &&\qquad \text{from \eqref{eq:pred-func}}
\end{align*}
\caption{A logic program limiting a predicate to be a function.}
\label{fig:pred-func}
\end{figure}

A solution to a CSP problem is then given by an interpretation of the
predicate $p_f/2$ in a model of the logic program obtained by
replacing the functions $f/1$ with the predicates $p_f/2$ and adding
the theory defining $p_f/2$. Different models give rise to different
solutions.%
Applying this transformation on the n-Queens problem, one obtains the
specification given in figure \ref{fig:lp-queens} where $pos(C,R)$ is
the predicate which gives the row $R$ of a queen at a column $C$.

\begin{figure}[hbt]
\begin{align*}
&safe(X_1,X_2,D)\la X_1\neq X_2, abs(X_1-X_2)\neq D. \\
&has\_pos(X)\la d_{row}(Y),pos(X,Y).
\end{align*}
\begin{align*}
has\_pos(X)&\La d_{col}(X). \\
d_{col}(X) &\La pos(X,Y).   \\
Y=Z &\La pos(X,Y),pos(X,Z).
\end{align*}
\begin{align*}
&safe(R_1,R_2,C_2-C_1) \La \\
&\qquad pos(C_1,R_1),pos(C_2,R_2),C_1<C_2.
\end{align*}
\caption{N-Queens as an (Abductive) Logic Program}
\label{fig:lp-queens}
\end{figure}

\subsection{Abduction}

Abductive logic programming \cite{abd:ALP} is a form of reasoning in
which an answer to a query is a set of facts. More formally, an
abductive framework is defined as a triple $\seq{P,A,I}$ where $P$ is
a logic program, $A$ is a set of predicates called {\em abducibles}
and $I$ is a set of integrity constraints. A solution is a set
$\Delta\subseteq A$\footnote{Here and in the rest of the paper we will
  use the same symbol $A$ to indicate both the set of abducible
  predicates and the set of all their ground instances.} such that
$M\models I$ for some canonical model $M$ of $P\cup\Delta$. If one is
interested in a query, then it can be put as part of the integrity
constraints. Different choices for the type of canonical model have
been considered in the literature. In \cite{abd:GSM}, $M$ must be a
stable model of $P\cup\Delta$ which is also called {\em generalized
  stable model} $M(A)$ of $P$. In \cite{abd:SLDNFA} three
valued models of the completion of the program are considered and the
abductive predicates must have a two-valued interpretation.

A problem defined as in figure \ref{fig:lp-queens} can be given
directly to an abductive procedure by declaring the predicate
$pos(X,Y)$ as abducible.

Proof procedures for abduction are defined in a similar way
\cite{abd:nf} to, or as an extension \cite{abd:SLDNFA} of SLDNF
resolution. In each state of the derivation they also maintain a set
$\Delta$ of already abduced atoms. When an abductive atom is selected
in the current goal, it is checked if it can be resolved with an atom
from $\Delta$. If this is not the case then the selected atom is added
to the set of abducibles and a consistency derivation is started which
checks the integrity constraints to see if this assumption does not
lead to a contradiction. If the abducible atoms contain variables
then during an abductive step these variables are replaced with skolem
constants and the unification algorithm is extended to deal with them
\cite{abd:SLDNFA}. We have already discussed the problems of the SLD
resolution for efficiently solving CSPs thus we can expect that the
performance of such more complex proof procedures will be even worse.

Recently, abduction has been extended to constraint logic programs.
One of the main ideas is that the skolem constants which are added as
arguments of the abduced predicates are in fact existentially
quantified constraint variables and their values can be computed by a
constraint solver. This allows us to have a more declarative
representation of CSPs and still use efficient techniques for
computing their solution. The first such integration is the ACLP
system of Kakas \cite{abd:ACLP@ICLP95} which is based on the proof
procedure of \cite{abd:nf}. Originally, ACLP was defined only for
definite programs and integrity constraints and in
\cite{abd:ACLP@JLP00} it was extended to deal with negation as failure
through abduction in a similar way as in \cite{abd:nf}. A more recent
integration of abduction and CLP is the SLDNFAC system
\cite{abd:SLDNFA-CLP} which is based on the abductive procedure SLDNFA
\cite{abd:SLDNFA}.

\subsection{Logic Programming with Stable Model Semantics}

Another way of using ``open'' predicates in a logic program is to use
the stable model semantics \cite{sem:SMS}. 
A predicate $p$ can be defined as having an open interpretation by
the following two rules:
\begin{align*}
  p         & \la \Not \ol{p}. \\
  \ol{p} & \la \Not p. 
\end{align*}
where $\ol{p}$ is a new predicate symbol. This program has two
stable models - $\{p\}$ and $\{\ol{p}\}$. If $A$ is a set of
predicates then we define with $T(A)$ the following set of clauses
\begin{align*}
T(A) = &\set{p(\vec{x})\la \Not \ol{p}(\vec{x}). | p\in A}\cup \\
       &\set{\ol{p}(\vec{x})\la \Not p(\vec{x}). | p\in A} 
\end{align*}
Then to compute a stable model of a CSP as represented in section
\ref{fig:lp-queens}, one only needs to add to the program the clauses
$T(A)$, where $A$ is the set of predicates which are the result of the
transformation of the function symbols\footnote{In fact, exactly the
  same methodology for representing CSPs is proposed in
  \cite{mg:smodels}}.

In fact, there is a very strong relation between the semantics of an
abductive framework and the stable models of an equivalent logic
program. It has been shown in \cite{abd:tms} that $M(\Delta)$ is a
generalized stable model of an abductive framework $\seq{P,A,I}$ iff
there exists a stable model $M'$ of $\seq{P\cup T(A),I}$ such that
$M'=M(\Delta)\cup\ol{\nabla}$ where
$\ol{\nabla}=\set{\ol{p}(\vec{x}) | p(\vec{x})\in A\backslash\Delta}$.

The rules declaring the open predicates can be combined with some of
the integrity constraints to obtain a more compact representation of
the problem.  For example, the integrity constraint stating that the
open predicate $p_f$ should be false for all values not in the domain
$d_p$
\[ d_p(X) \La p_f(X,Y)  \]
can be omitted by using the rule
\[ p_f(X,Y)\la d_p(X), \Not \ol{p_f}(X,Y).\] 
The reason is that in the stable model semantics an atom can be true
only if it is a head of some clause. The full specification of the
n-Queens problem is given in figure \ref{fig:sm-queens}.

\begin{figure}[hbt]
\begin{program}
\ind pos(X,Y)\la d_{col}(X), \Not \ol{pos}(X,Y).
\ol{pos}(X,Y)\la \Not pos(X,Y).
~
safe(X_1,X_2,D)\la X_1\neq X_2, abs(X_1-X_2)\neq D.
has\_pos(X)\la d_{row}(Y),pos(X,Y).
~
has\_pos(X)\La d_{col}(X).
Y=Z \La pos(X,Y),pos(X,Z).
~
safe(R_1,R_2,C_2-C_1)\La
\ind pos(C_1,R_1),pos(C_2,R_2),C_1<C_2.
\end{program}
\caption{N-Queens with Stable Models Semantics}
\label{fig:sm-queens}
\end{figure}

An efficient implementation of the stable model semantics is the
\smodels\ system \cite{mg:NS96}. It works with propositional rules and
a special pre-processing program is used for grounding function-free
range-restricted logic programs. The implementation of the system
combines bottom-up inference with backtracking search and employs
powerful pruning methods.

\subsection{Other Formalisms}

In \cite{sem:data-circ} is presented a language called \datalogc\ 
which is an extension of \datalog\ where only some of the predicates
are minimized and the interpretation of the others is left open. The
semantics of the language originates from the nonmonotonic formalism
of {\em circumscription} and is defined as the minimal Herbrand model
of the program w.r.t.\ a fixed interpretation of the open predicates.
It is proven in \cite{sem:data-circ} that the data complexity of
deciding whether a query is not entailed by the program is NP-complete
which means that one can express any CSP in this formalism. However,
as the language does not contain negation, one can not use directly
the methodology discussed above for representing CSPs.

In \cite{np-spec} is defined the language \npspec\ which is an
extension of \datalogc\ and allows a more natural representation of
problems. The main difference is that it supports special
meta-declarations, called {\em tailoring predicates}, restricting the
domain and the interpretation of the open predicates. The most simple
one is of the form $Subset(domain/n, pred)$ which defines $pred/n$ to
be an open predicate and its interpretation should be a subset of the
interpretation of the predicate $domain/n$. Another declaration
is $Partition(domain/n,pred,m)$ which states that the predicate
$pred/n$ should be partitioned in $m$ sets which is, in fact, equivalent
to a function with domain $domain/n$ and a range $0..m-1$.
The two other tailoring predicates are $Permutation(domain/n,pred)$
and $IntFunc(domain/n,pred,min..max)$ which express respectively that
$pred/n$ is a bijection from $domain/n$ to $0..|domain/n|-1$ and
$pred/n$ is a function with a range $min..max$. Another extension of
\npspec\ is the support of a predefined arithmetic functions and
predicates.

The formulation of the graph coloring problem in the \npspec\ language
is given below. The input of the program consists of facts $node/1$
and $edge/2$ describing the graph.
\begin{align*}
&Partition(node/1,color,4). \\[2pt]
&\La edge(V_1,V_2),color(V_1,C_1),color(V_2,C_2).
\end{align*}

\section{Experiments}

\subsection{The Systems}

The finite domain CLP package is the one provided with SICStus version
3.7. Given the reputation of SICStus and of finite domain CLP, one can
assume it offers state of the art technology for CSP solving and it is
a good yardstick to judge the performance of other systems. The
abductive system ACLP \cite{abd:ACLP@JLP00} is a meta interpreter
written in Prolog, runs on Eclipse version 4.2 and makes use of its
finite domain package. The abductive system SLDNFAC
\cite{abd:SLDNFA-CLP} is also a meta interpreter written in Prolog but
runs on SICStus version 3.7 and makes use of the SICStus finite domain
package. The model generator SEM \cite{mg:SEM} version 1.7 is a fine
tuned package written in C. \smodels\ \cite{mg:NS96} version 2.25, the
system for computing stable models is implemented in C++ and the
associated program used grounding is {\it lparse} version 0.99.48. All
experiments have been done on the same hardware, namely Pentium II.

All systems based on a finite domain constraint solver used a labeling
strategy which first selects variables with the smallest domain and
then the ones which participate in the highest number of
constraints\footnote{This strategy is sometimes abbreviated to $|ffc|$.}.

\subsection{N-Queens}

\begin{figure*}[p]
  \begin{center}
  \psfig{file=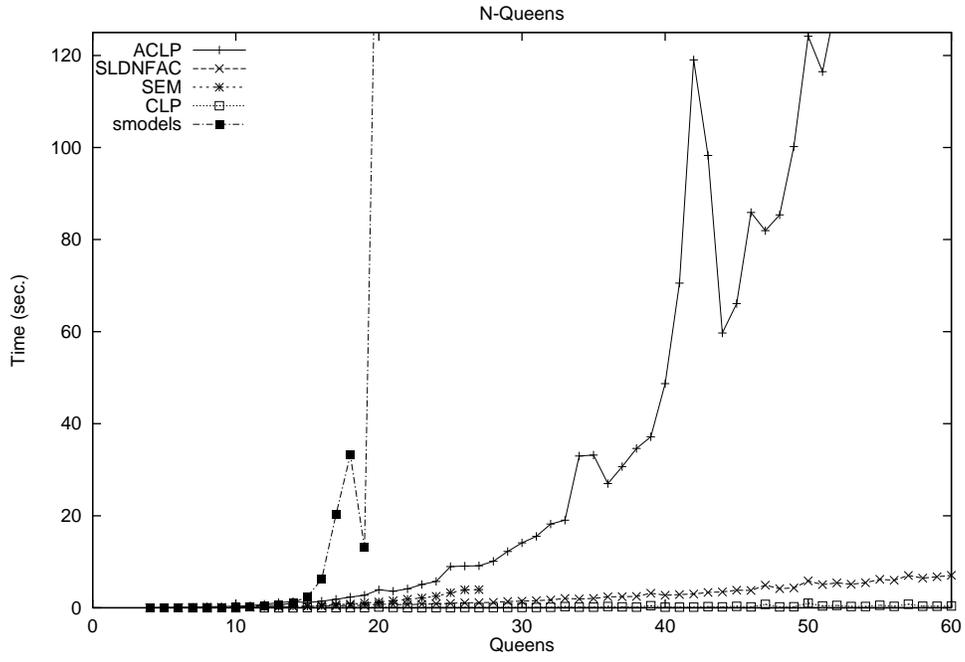}
  \end{center}
  \caption{N-Queens: Running times} \label{fig:q}
\end{figure*}

\begin{figure*}[p]
  \begin{center}
  \psfig{file=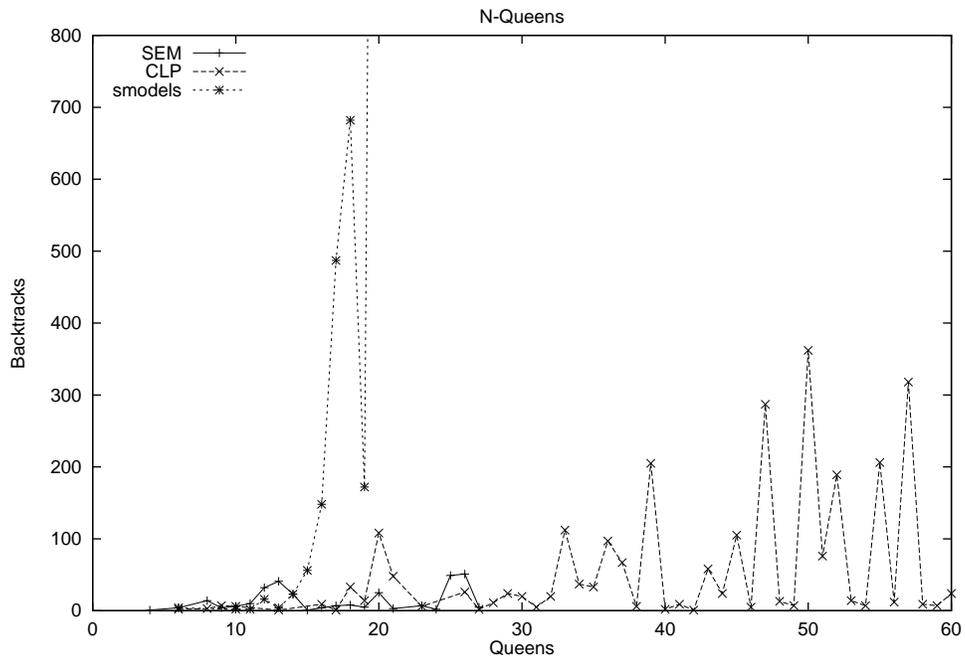}
  \end{center}
  \caption{N-Queens: Number of backtracks.} \label{fig:back}
\end{figure*}

Figure \ref{fig:q} gives the running times for the different systems
and figure \ref{fig:back} gives the number of backtracks. The two
abductive systems (ACLP and SLDNFAC) do not introduce any extra choice
points compared to CLP and hence are not plotted in figure
\ref{fig:back}.
Not surprisingly, CLP gives the best results.  SLDNFAC is second best
and, although meta-interpretation overhead increases with problem
size, deteriorates very slowly.  SEM is third but runs out of memory
for large problems (it needs about 120MB for 27 queens).  This is
probably caused by a not very good techniques for grounding the
problem and exploring the search space.  The times given for SEM do
not include time spend by the operating system in managing the memory
which becomes considerable for the larger instances of the problem.
ACLP performs substantially worse than SLDNFAC and degrades more
quickly for the larger problems. It can likely be attributed to the
more experimental nature of the implementation. \smodels\ performs very
poorly on this problem, in particular when compared with its
performance on the graph coloring problem. 
As can be seen from figure \ref{fig:back} the main reason seems to be
the large number of backtracks it does.

The CLP consistency techniques seem to be much less sensitive to the
domains size, and this carries over to the abductive systems which
reduce the problem to a CLP problem and then use the CLP solver to
search for the solution.

\subsection{Graph Coloring}

\begin{figure*}[htbp]
\begin{center}
  \psfig{file=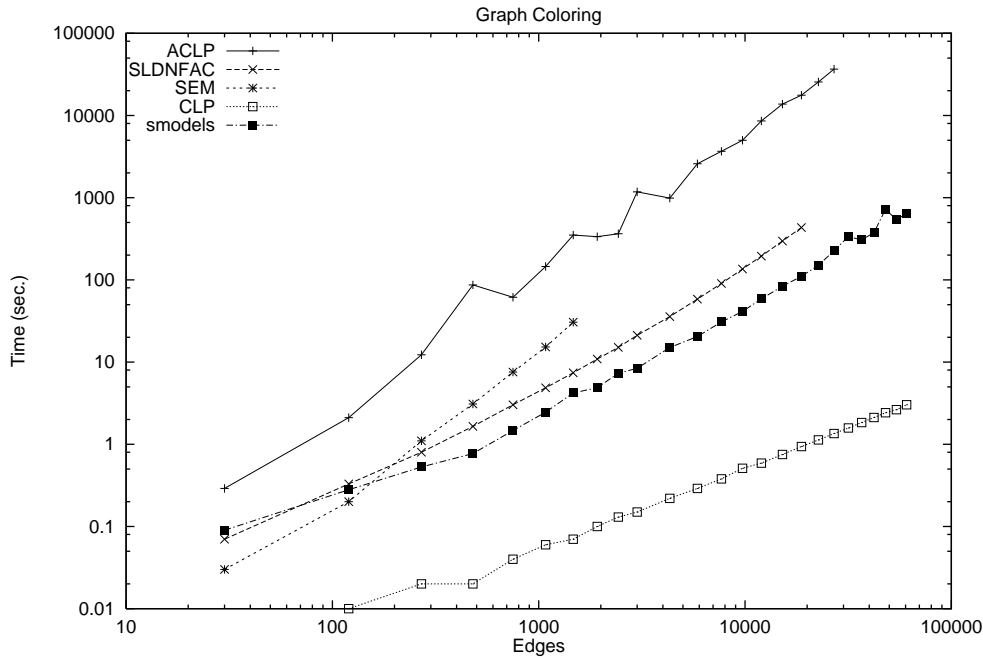}
\end{center}
\caption{Graph Coloring} \label{fig:gc}
\end{figure*}

We used a graph generator program which is available from address
\url{http://web.cs.ualberta.ca/~joe/Coloring/Generators/generate.html}.
We did our experiments with planar undirected graphs which are known
to be 4-colorable. The graphs were generated using a 20\% probability
of assigning arcs. This results in dense graphs with a stable
behavior. For this problem, the domain of the solution variables (the
number of colors) remained the same and we have modified only the
parameter of the problem (the number of vertices) and consequently the
number of constraints (arcs). Figure \ref{fig:gc} gives the results of
solving the problem with the different systems.  Both axes are plotted
in a logarithmic scale. On the x-axis we have put the number of arcs
(constraints) instead of the number of vertices.

Not surprisingly, CLP is the fastest system. \smodels\ is second best
on this problem. We assume it is in part because of the very concise
formulation. Using the so called technique of rules with exceptions
\cite{mg:smodels}, the two rules needed to describe the space of
candidate solutions also encode the constraint that the color is a
function of the vertex.  Hence there is only one other rule, namely
the constraint that two adjacent vertices must have a different color.
The difference with CLP is almost two orders of magnitude for the
largest problems. SLDNFAC is slightly worse than \smodels. Although
meta-interpretation overhead tends to increase with problems size, the
difference with \smodels\ grows very slowly. The model generator SEM
deteriorates much faster and runs out of memory for the larger
problems. The fact that it grounds the whole theory is a likely
explanation. The difference with \smodels\ supports the claim that
\smodels\ has better techniques for grounding. ACLP performs
substantially worse than SLDNFA and also deteriorates faster.

\section{Conclusion}

The examples which we have considered in this paper are by no ways
representative. However we think that they still show some interesting
features and limitations of the considered systems.

Consistency algorithms are a very efficient way for solving CSP.
Constraint logic programming allows this techniques to be integrated
in a natural and clear way to logic programs. However, as argued
parameterized CSPs can not be represented in a declarative way as
CLP(FD) programs. Using abduction or stable models as the basis for
logic programming allows the problems to be represented in a more
declarative way and the recent integration of abduction with CLP
allows the same consistency techniques to be used for solving the
problems. At the moment, such systems are implemented as
meta-interpreters on top of Prolog and they essentially reduce a
problem to the same set of constraints (in many cases without
backtracking) which would be produced by the corresponding constraint
logic program. Our experiments suggest that the overhead of an
abductive system is small and acceptable. Moreover they can also solve
other classes of problems which require non-monotonic reasoning like
planning problems. We also showed that there is a very close relation
between the semantics and the problem representation of abduction and
logic programming with stable model semantics. The only difference is
in the reasoning techniques - abduction is usually done by a top-down
proof procedure, while a stable model is usually computed by a
bottom-up procedure.  However, the techniques used to compute a stable
model of a program do not seem to be so well suited for solving CSPs.
One reason could be that they work on a ground propositional programs
which tend to be large and grow fast as the parameter of the problem
increases. This suggests that an interesting area for further research
would be a framework for computing stable models of constraint logic
programs with the help of constraint solving techniques. Some work has
already been done in this direction \cite{sem:nm-clp,sem:eiter}.

We argued earlier in the paper that the most natural way for
representing CSPs is with functions with open interpretation. Hence it
would be interesting to consider extensions of the CLP scheme which
directly support such form of reasoning. Some work in this area is
done in \cite{clp:of,clp:Hickey93}.

\section{Acknowledgments.}

We want to thank the members of the DTAI group at K.U.Leuven and
anonymous referees for their useful comments. This research is
supported by the GOA project LP+. The third author is supported by
FWO-Vlaanderen.


\begin{thebibliography}{}

\bibitem[\protect\citeauthoryear{Bruynooghe, Pelov, \& Denecker}{1999}]{clp:of}
Bruynooghe, M.; Pelov, N.; and Denecker, M.
\newblock 1999.
\newblock Towards a more declarative language for solving finite domain
  problems.
\newblock In Apt, K.; Kakas, A.; Monfroy, E.; and Rossi, F., eds., {\em
  Proceedings of the ERCIM/COMPULOG Workshop on Constraints}.
\newblock Paphos, Cyprus: University of Cyprus.

\bibitem[\protect\citeauthoryear{Bruynooghe}{1991}]{ib:mb-91}
Bruynooghe, M.
\newblock 1991.
\newblock Intelligent backtracking revisited.
\newblock In Lassez, J.-L., and Plotkin, G., eds., {\em Computational Logic -
  Essays in Honor of Alan Robinson}. MIT Press.
\newblock  166--177.

\bibitem[\protect\citeauthoryear{Cadoli \& Palopoli}{1998}]{sem:data-circ}
Cadoli, M., and Palopoli, L.
\newblock 1998.
\newblock Circumscribing \textsc{datalog}: expressive power and complexity.
\newblock {\em Theoretical Computer Science} 193(1--2):215--244.

\bibitem[\protect\citeauthoryear{Cadoli \bgroup \em et al.\egroup
  }{1999}]{np-spec}
Cadoli, M.; Palopoli, L.; Schaerf, A.; and Vasile, D.
\newblock 1999.
\newblock \textsc{np-spec}: An executable specification language for solving
  all problems in {NP}.
\newblock In Gupta, G., ed., {\em 1st International Workshop on Practical
  Aspects of Declarative Languages}, volume 1551 of {\em Lecture Notes in
  Computer Science},  16--30.
\newblock San Antonio, Texas, USA: Springer-Verlag.

\bibitem[\protect\citeauthoryear{Codognet \& Diaz}{1996}]{clp:clp(FD)}
Codognet, P., and Diaz, D.
\newblock 1996.
\newblock Compiling constraints in clp({FD}).
\newblock {\em Journal of Logic Programming} 27(3):185--226.

\bibitem[\protect\citeauthoryear{Denecker \& De~Schreye}{1998}]{abd:SLDNFA}
Denecker, M., and De~Schreye, D.
\newblock 1998.
\newblock {SLDNFA}: an abductive procedure for abductive logic programs.
\newblock {\em Journal of Logic Programming} 34(2):201--226.

\bibitem[\protect\citeauthoryear{Denecker \&
  Van~Nuffelen}{1999}]{abd:SLDNFA-CLP}
Denecker, M., and Van~Nuffelen, B.
\newblock 1999.
\newblock Experiments for integration {CLP} and abduction.
\newblock In Apt, K.; Kakas, A.; Monfroy, E.; and Rossi, F., eds., {\em
  Proceedings of the 1999 ERCIM/COMPULOG Workshop on Constraints}.
\newblock Paphos, Cyprus: University of Cyprus.

\bibitem[\protect\citeauthoryear{Dix \& Stolzenburg}{1998}]{sem:nm-clp}
Dix, J., and Stolzenburg, F.
\newblock 1998.
\newblock A framework to incorporate non-monotonic reasoning into constraint
  logic programming.
\newblock {\em Journal of Logic Programming} 37(1--3):47--76.

\bibitem[\protect\citeauthoryear{Eiter, Lu, \& Subrahmanian}{1997}]{sem:eiter}
Eiter, T.; Lu, J.~J.; and Subrahmanian, V.~S.
\newblock 1997.
\newblock Computing non-ground representations of stable models.
\newblock In Dix, J.; Furbach, U.; and Nerode, A., eds., {\em Proceedings of
  the Fourth International Conference on Logic Programming and Non-Monotonic
  Reasoning}, volume 1265 of {\em Lecture Nortes in Computer Science},
  198--217.
\newblock Dagstuhl, Germany.

\bibitem[\protect\citeauthoryear{Eshghi \& Kowalski}{1989}]{abd:nf}
Eshghi, K., and Kowalski, R.
\newblock 1989.
\newblock Abduction compared with negation by failure.
\newblock In Levi, G., and Martelli, M., eds., {\em Proceedings of the Sixth
  International Conference on Logic Programming},  234--254.
\newblock Lisbon, Portugal.

\bibitem[\protect\citeauthoryear{Gelfond \& Lifschitz}{1988}]{sem:SMS}
Gelfond, M., and Lifschitz, V.
\newblock 1988.
\newblock The stable model semantics for logic programming.
\newblock In Kowalski, R.~A., and Bowen, K.~A., eds., {\em Logic Programming,
  Proceedings of the Fifth International Conference and Symposium},
  1070--1080.
\newblock Seattle, Washington: MIT Press.

\bibitem[\protect\citeauthoryear{Hickey}{1993}]{clp:Hickey93}
Hickey, T.~J.
\newblock 1993.
\newblock Functional constraints in {CLP} languages.
\newblock In Benhamou, F., and Colmerauer, A., eds., {\em Constraint Logic
  Programming: Selected Research}. MIT Press.
\newblock  355--381.

\bibitem[\protect\citeauthoryear{Jaffar \& Maher}{1994}]{clp:survey@JLP}
Jaffar, J., and Maher, M.
\newblock 1994.
\newblock Constraint logic programming: A survey.
\newblock {\em Journal of Logic Programming} 19/20:503--581.

\bibitem[\protect\citeauthoryear{Kakas \& Mancarella}{1990}]{abd:GSM}
Kakas, A.~C., and Mancarella, P.
\newblock 1990.
\newblock Generalized stable models: A semantics for abduction.
\newblock In {\em Proceedings of the 9th ECAI},  385--391.

\bibitem[\protect\citeauthoryear{Kakas \& Michael}{1995}]{abd:ACLP@ICLP95}
Kakas, A.~C., and Michael, A.
\newblock 1995.
\newblock Integrating abductive and constraint logic programming.
\newblock In Sterling, L., ed., {\em Proceedings of the 12th International
  Conference on Logic Programming},  399--413.
\newblock Tokyo, Japan.

\bibitem[\protect\citeauthoryear{Kakas, Kowalski, \& Toni}{1992}]{abd:ALP}
Kakas, A.~C.; Kowalski, R.; and Toni, F.
\newblock 1992.
\newblock Abductive logic programming.
\newblock {\em Journal of Logic and Computation} 2(6):719--770.

\bibitem[\protect\citeauthoryear{Kakas, Michael, \&
  Mourlas}{2000}]{abd:ACLP@JLP00}
Kakas, A.~C.; Michael, A.; and Mourlas, C.
\newblock 2000.
\newblock {ACLP}: Abductive constraint logic programming.
\newblock {\em Journal of Logic Programming}.
\newblock To appear.

\bibitem[\protect\citeauthoryear{Lloyd}{1987}]{lloyd}
Lloyd, J.~W.
\newblock 1987.
\newblock {\em Foundations of Logic Programming}.
\newblock Springer-Verlag, second edition.

\bibitem[\protect\citeauthoryear{Mackworth}{1992}]{clp:Mackw92}
Mackworth, A.~K.
\newblock 1992.
\newblock The logic of constraint satisfaction.
\newblock {\em Journal of Artificial Intelligence} 58(1--3):3--20.

\bibitem[\protect\citeauthoryear{Niemel\"a \& Simons}{1996}]{mg:NS96}
Niemel\"a, I., and Simons, P.
\newblock 1996.
\newblock Efficient implementation of the well-founded and stable model
  semantics.
\newblock In Maher, M., ed., {\em Logic Programming, Proceedings of the 1996
  Joint International Conference and Syposium},  289--303.
\newblock Bonn, Germany: MIT Press.

\bibitem[\protect\citeauthoryear{Niemel{\"a}}{1999}]{mg:smodels}
Niemel{\"a}, I.
\newblock 1999.
\newblock Logic programs with stable model semantics as a constraint
  programming paradigm.
\newblock {\em Annals of Mathematics and Artificial Intelligence}
  25(3,4):241--273.

\bibitem[\protect\citeauthoryear{Satoh \& Iwayama}{1991}]{abd:tms}
Satoh, K., and Iwayama, N.
\newblock 1991.
\newblock Computing abduction by using the {TMS}.
\newblock In Furukawa, K., ed., {\em Proceedings of the Eighth International
  Conference on Logic Programming},  505--518.
\newblock Paris, France.

\bibitem[\protect\citeauthoryear{Slaney}{1995}]{mg:finder}
Slaney, J.
\newblock 1995.
\newblock {FINDER} version 3.0 - notes and guides.
\newblock Technical report, Centre for Information Science Research, Australian
  National University.

\bibitem[\protect\citeauthoryear{Tsang}{1993}]{clp:Tsang}
Tsang, E.
\newblock 1993.
\newblock {\em Foundations of Constraint Satisfaction}.
\newblock Computation in Cognitive Science. Academic Press.

\bibitem[\protect\citeauthoryear{Van~Hentenryck, Saraswat, \&
  Deville}{1998}]{clp:cc(FD)}
Van~Hentenryck, P.; Saraswat, V.~A.; and Deville, Y.
\newblock 1998.
\newblock Design, implementation, and evaluation of the constraint language
  cc({FD}).
\newblock {\em Journal of Logic Programming} 37(1--3):139--164.

\bibitem[\protect\citeauthoryear{Van~Hentenryck}{1989}]{clp:cslp}
Van~Hentenryck, P.
\newblock 1989.
\newblock {\em Constraint Satisfaction in Logic Programming}.
\newblock MIT Press.

\bibitem[\protect\citeauthoryear{Zhang \& Zhang}{1995}]{mg:SEM}
Zhang, J., and Zhang, H.
\newblock 1995.
\newblock {SEM}: a system for enumerating models.
\newblock In Mellish, C.~S., ed., {\em Proceedings of the Fourteenth
  International Joint Conference on Artificial Intelligence},  298--303.
\newblock San Mateo: Morgan Kaufmann.

\end{thebibliography}
\end{document}